\documentclass {aa}  
\pdfoutput=1
\usepackage{graphicx}
\usepackage{txfonts}
\usepackage{amsmath}
\usepackage{subfigure}
\usepackage{float}
\usepackage{xcolor}

\urldef{\stixads}\url{https://ui.adsabs.harvard.edu/abs/2020A%26A...642A..15K/abstract}

\begin{document}

   \title{Characterising fast-time variations in the hard X-ray time profiles of solar flares using Solar Orbiter's STIX}


   \author{Hannah Collier \inst{1}\fnmsep\inst{2}\thanks{E-mail: hannah.collier@fhnw.ch}
          \and 
          Laura A. Hayes \inst{3}
          \and
          Andrea F. Battaglia \inst{1}\fnmsep\inst{2}
          \and
          Louise K. Harra \inst{2}\fnmsep\inst{4}
          \and
          S\"am Krucker \inst{1}\fnmsep\inst{5}
          }

        \institute{University of Applied Sciences and Arts Northwestern Switzerland, Bahnhofstrasse 6, 5210 Windisch, Switzerland
        \and
            ETH Z\"{u}rich,
              R\"{a}mistrasse 101, 8092 Z\"{u}rich Switzerland
        \and
            European Space Agency, ESTEC,
            Keplerlaan 1 - 2201 AZ, Noordwijk, The Netherlands
        \and
             PMOD/WRC, Dorfstrasse 33, CH-7260 Davos Dorf, Switzerland
        \and
            Space Sciences Laboratory, University of California, 7 Gauss Way, 94720 Berkeley, USA
            }

   \date{Received October 26, 2022; accepted December 26, 2022}

 
  \abstract
   {}
   {The aim of this work is to develop a method to systematically detect and characterise fast-time variations ($\gtrsim 1$s) in the non-thermal hard X-ray (HXR) time profiles of solar flares using high-resolution data from Solar Orbiter's Spectrometer/Telescope for Imaging X-rays (STIX).}
   {The HXR time profiles were smoothed using Gaussian Process (GP) regression. The time profiles were then fitted with a linear combination of Gaussians to decompose the time profile. From the Gaussian decomposition, key characteristics such as the periodicity, full width at half maximum (FWHM), time evolution, and amplitude can be derived.}
   {We present the outcome of applying this method to four M and X GOES-class flares from the first year of Solar Orbiter science operations. The HXR time profiles of these flares were decomposed into individual Gaussians and their periods were derived. The quality of fit is quantified by the standard deviation of the residuals (difference between observed and fitted curve, normalised by the error on the observed data), for which we obtain $\leq 1.8$ for all flares presented. In this work, the first detection of fast-time variations with Solar Orbiter's STIX instrument has been made on timescales across the range of 4-128s.} 
    {A new method for identifying and characterising fast-time variations in the non-thermal HXR profiles of solar flares has been developed, in which the time profiles are fit with a linear combination of Gaussian bursts. The opportunity to study time variations in flares has greatly improved with the new observations from STIX on Solar Orbiter.}

\titlerunning{Characterising fast-time variations in HXR flares}
\authorrunning{H. Collier et al}

\maketitle
%

\section{Introduction}
Solar Orbiter is a solar and heliospheric mission led by the European Space Agency (ESA) in partnership with NASA. The scientific payload of Solar Orbiter includes four in situ and six remote sensing instruments, including the Spectrometer/Telescope for Imaging X-rays (STIX). STIX is a hard X-ray imaging spectrometer that detects photons with energies in the range of 4-150 keV, with a 1 keV energy resolution (at 6 keV) \citep{Krucker_2020}.  STIX measures bremsstrahlung emission from solar flares and therefore provides diagnostics on the hottest ($\gtrsim8$ MK) flare plasma \citep{Krucker_2020,2021A&A...656A...4B}. This means that STIX is equipped to provide information on the accelerated electrons producing such bremsstrahlung emissions upon Coulomb collisions with ambient ions. STIX has a high time resolution, with the ability to sample down to 0.1 s. It also has a stable background and continuously observes the full solar disc. Thanks to these capabilities, STIX is an excellent instrument for measuring time-varying signatures on short timescales ($\gtrsim1$s) in the X-ray emission of solar flares. 
\\
\\
Such time variations are of great interest because they are related to the fundamental timescales occurring in solar flares, such as energy release and particle acceleration processes, as well as magnetohydrodynamic (MHD) waves and oscillations in or around the flare site. It is imperative to understand the origin and nature of the observed time-varying behaviour in order to achieve a unified solar flare model.
\\
\\
Fast time variations, which are sometimes classified as quasi-periodic pulsations (QPPs), have been observed in the emission from solar flares over the past 50 years, with some of the earliest studies identifying time-varying signatures in the hard X-ray (HXR) energy range \citep{parksandwinckler}. The particular variations of interest in this work are modulations in the flare intensity-time profiles. The ones classified as QPPs typically have periodicities ranging from a few seconds to several minutes \citep{zimovets2021}. Sub-second spikes in the HXR flare emission have also been detected \citep{Roberts1983,rhessi_subsec_casestudy,Knuth_2020}. In some cases, multiple periods of oscillations have been identified alongside amplitude modulation \citep{2015A&A...574A..53K,vandoorsselaere_2016,mclaughlin2018}. Furthermore, \citet{Hayes_2016} identified these time-varying signatures in both the impulsive and decay phases, thus challenging our understanding of the flare model. Additionally, studies relating active region and flare properties to QPP periodicities have also been performed \citep{pugh2017, Hayes_2020}. For example, \cite{Hayes_2020} found a positive correlation between QPP period and duration. Despite this, QPP period was found to be independent of flare magnitude. Finally, a significant correlation between QPP period and various ribbon properties was found. 
\\
\\
Statistical studies have found a wide range of probabilities that a given $>M5$ GOES class flare will contain a QPP, with estimates ranging from 30-90\% \citep{simoes2015, inglis2017, dominique2018,  Hayes_2020}. It has been made clear that time-varying signatures (whether quasi-periodic or not) are commonly observed phenomena which occur in flare emission across the entire electromagnetic spectrum, from radio waves to $\gamma$- rays \citep{nakariakov2009, zimovets2021}. These statistics rely on QPP classifications that search for a statistically significant periodic component in the data and thus exclude all fast-time-varying structures that do not display any significant periodicity. Recent reviews of QPP properties include \citet{zimovets2021, Kupriyanova2020quasi-periodic,mclaughlin2018,  vandoorsselaere_2016, nakariakov2009}. 
\\
\\
A number of models have been built to attempt to explain the observed time-varying and oscillatory phenomena in solar flare emission, with at least fifteen different   models proposed at present \citep{zimovets2021}. For a recent review of current models and their observational signatures, we refer to \citet{zimovets2021} and \citet{mclaughlin2018}. According to \cite{Kupriyanova2020quasi-periodic}, these models can generally be divided into three categories: (1) models in which the observed emission is directly modulated by MHD and electromagnetic (EM) waves; (2) models in which the efficiency of energy release is modulated by MHD waves; and (3) models in which the original energy release process is itself quasi-periodic in nature. Despite the many observations of time-varying behaviour in flares, there are few cases where the underlying mechanism(s) behind the variations are unambiguous. This is often due to observational constraints. In addition to this, many mechanisms have non-unique signatures, making the disambiguation more challenging \citep{mclaughlin2018, zimovets2021}.
\\
\\
One of the challenges in the study of time-varying signatures and QPPs in solar flares is the identification of the characteristic timescale or period. \cite{broomhall} analysed the strengths and weaknesses of various state-of-the-art techniques for detecting time-varying signatures classified as QPPs, based on a series of tests. The tests involved a simulated dataset of flare time profiles, some of which include a periodic component. From this analysis, a list of eight recommendations for detection methods was set out. One of these recommendations relates to a common pre-processing step, namely: time-series detrending. \citet{broomhall} showed that detrending can be challenging and if it must be done, it should be performed manually, with a time-dependent smoothing window; otherwise, a bias may be introduced. Furthermore, there is the added complexity of the underlying background trend when analysing coronal time series. \citet{Auch_re_2016} showed that if the power-law dependency of the Fourier power spectra (also known as red noise) is not considered in detection models, this can lead to false detections of a significant periodic component. \cite{Inglis_2015} showed the effect of this power-law component in the Fourier power spectrum and took this red noise component into consideration when detecting QPPs with methods such as AFINO \citep{inglis2017}. Many time-varying signatures in flares also show non-stationary properties, whereby the amplitude and period change as a function of time \citep{nak2019}. This is a further challenge to detection methods, for which typical Fourier-based approaches do not work well.
\\
\\
The method developed in this work leverages the recent data from Solar Orbiter's STIX and takes a new approach to analysing fast-time-varying signatures in solar flare HXR time profiles. Since the modulation depth in the HXR signature is large, we assume that detrending is not a necessary step. Instead, the fast-time-varying behaviour is considered here as a linear combination of individual Gaussian contributions to the total observed signal. To achieve this, first the HXR time series are smoothed using Gaussian Process (GP) Regression. Next, the time profiles are fitted with a linear combination of Gaussians. From the Gaussian decomposition, key characteristics such as the waiting time distribution, time evolution of the peak full width at half maximum (FWHM), and amplitude can be derived. Additionally, this method can detect non-stationary (time-evolving) signatures as the method is agnostic to whether the underlying driver is periodic. The derived timing information can be used to spatially resolve sources and to perform time-dependent spectral analyses of the individual peaks. An important distinction between this method and those mentioned in this section is that the method does not rely on the assumption that the observed oscillatory behaviour is periodic. This method can be applied to HXR time profiles which exhibit fast-time variations that may or may not be identified as quasi-periodic by existing analysis techniques. This aspect broadens the scope of our analysis.
\\
\\
The analysis given in this work is based on four M and X GOES-class flares observed by STIX as of September 2021. The opportunity to study time variations in flares has greatly improved thanks to the new observations from STIX on Solar Orbiter, which include hundreds of flares that demonstrate a fast-time variability. 
\section{Observations}
The data set used in this work is from the STIX imaging spectrometer onboard Solar Orbiter. STIX creates images by an indirect Fourier based imaging technique \citep{Krucker_2020}. It has a movable attenuator which is inserted during large flares to limit exposure to high count rates from low-energy X-ray photons and to avoid instrument saturation \citep{Krucker_2020}. STIX is capable of quantifying the location, spectrum, and energy content of flare-accelerated thermal and non-thermal electrons \citep{Krucker_2020}. 
\\
\\
STIX has an onboard algorithm which performs the dynamic time binning. This means that the time resolution of data taken by STIX can vary throughout the duration of the flare. The highest possible time cadence is 0.1 s. To date, the highest cadence tested is 0.3 s. It is not possible to take data at 0.1s for a long time ($\gtrsim$ 1 hour) since this fills up the onboard memory. However, it is possible to run at 0.5s cadence for as long as desired. As such, the time resolution of the data used for analysis in this work is 0.5s. The overview plots shown in this work are made using lower latency data, with a dynamically-adjusted temporal resolution, which typically ranges from 2 to 20 s. The energy ranges used vary depending on the individual flare profile and are selected such that they contain mainly non-thermal HXR emission. 
\\
\\
STIX has a relatively stable non-solar background and is a full Sun imager. It observes the Sun continuously, unlike its predecessor, the Reuven Ramaty High Energy Solar Spectroscopic Imager (RHESSI), which rotated about its own axis with a period of 4s \citep{Lin2002SoPh..210....3L}. This rotational period sometimes introduced artificial periodicities in the data, which were difficult to disentangle from true flare fluctuations \citep{inglis2011}. Due to its orbit, RHESSI had a day-night cycle, which meant that certain sections of a flare were often not observed. These constraints limited RHESSI's capability for analysing time-varying structures on short timescales. In contrast, STIX is an excellent instrument suited to the task of fast-time variations analysis due to its stable background, high temporal resolution, imaging and spectral capabilities. Furthermore, STIX is on board the Solar Orbiter spacecraft, which is in a unique elliptical orbit about the Sun,  observing solar dynamics from a different vantage point than Earth.

\section{Methodology}

\begin{figure*}
    \centering
    \includegraphics[width=\textwidth]{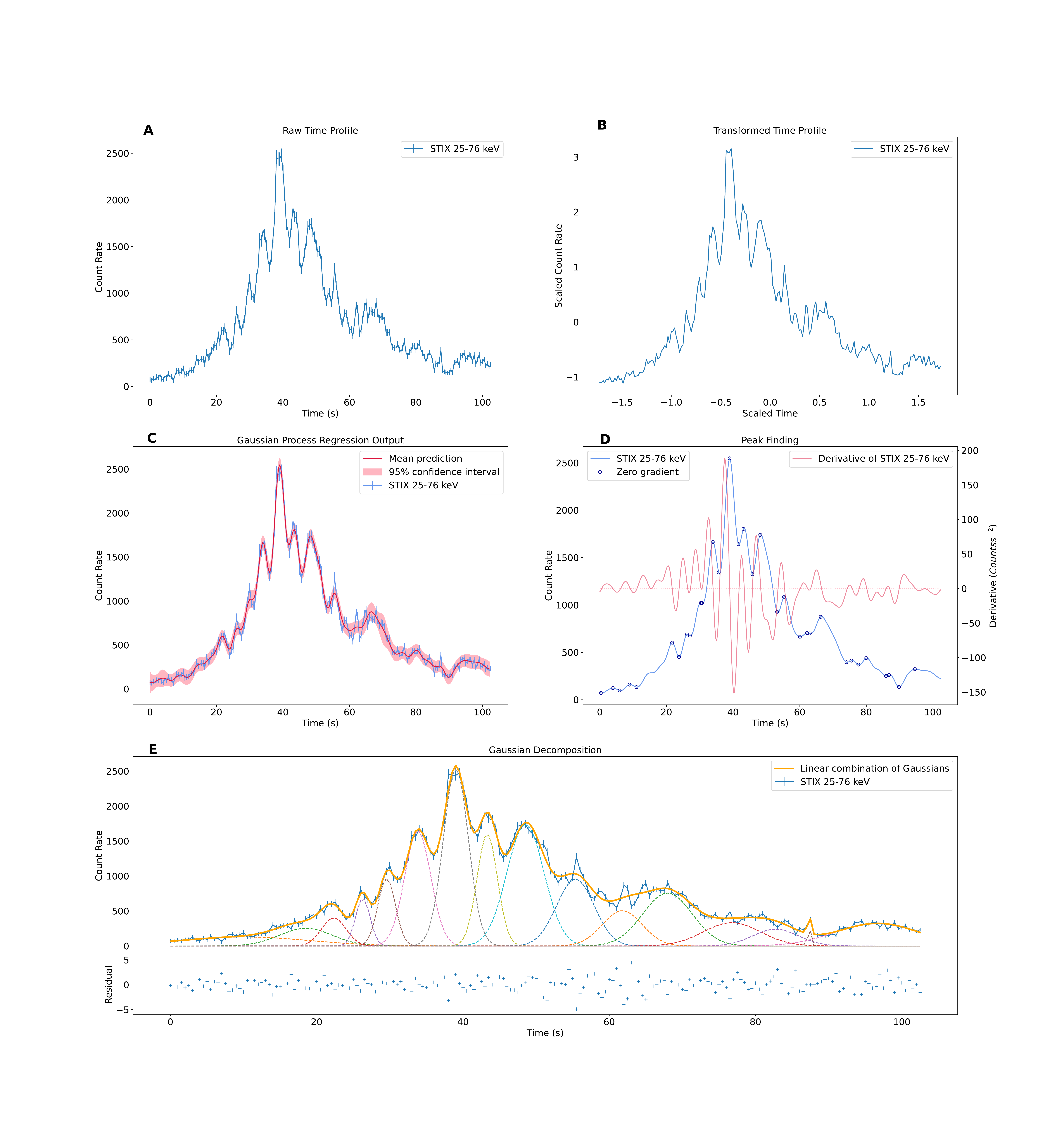}
    \caption{Various steps involved in the application of this method for the example M1.2 GOES class flare of SOL2022-05-04. The times shown are in seconds from 2022-05-04 15:16:12 (Earth time).}
    \label{fig:method_overview_plot}
\end{figure*}

The motivation for this work is to develop a method that  systematically identifies and characterises the modulation observed on short timescales in HXR non-thermal emission from solar flares. This characterisation is useful because it enables us to gain information about the timing, shape, and origin of HXR pulsations in a systematic way. The methodology involves two main steps: 1) data pre-processing by means of time series normalisation and signal smoothing and 2) Fitting a linear combination of Gaussians to the pre-processed time series. Each step and its motivation is explained in more detail in the following subsections\footnote{The code used here is publicly available at \url{https://github.com/hannahc243/Gaussian_Decomp}.}. 
\subsection{Data pre-processing}\label{subsection:preprocessing}
In this work, a Gaussian process (GP) regression is used to smooth the HXR time profiles. This  is a non-parametric, Bayesian approach to regression that uses machine learning techniques to fit the data. 
\subsubsection{Normalization}
To perform GP regression, the first step is normalising the data. To do so, we re-scale the original time profile using scikit-learn's StandardScalar class \citep{scikit-learn} by subtracting the mean, $\mu$ and dividng by the standard deviation, $\sigma,$ such that the new time profile is characterised by $\mu' = 0$ and $\sigma' = 1$.  
Figure \ref{fig:method_overview_plot} (A and B) demonstrates this for the STIX 25-76 keV time series of the SOL2022-05-04 flare. 

\subsubsection{Gaussian process (GP) regression}\label{subsection:GP_REG}
Gaussian processes (GPs) are a powerful supervised machine learning tool that provide a means to make predictions about data by incorporating prior knowledge. The most obvious area of application is regression problems, such as time-series forecasting. Overall, GPs are not limited to regression and can also be extended to classification and clustering tasks.
In this work, GP regression is used to smooth STIX time profiles. This is necessary  to identify the local maxima and minima for fitting. Since GP regression is a non-parametric approach, there is no initial assumption about the functional form of the time profile. In contrast, when performing polynomial regression, we assume that the functional form is a polynomial and then estimates the exact coefficients. Furthermore, GP regression uses Bayesian statistics, meaning that the model begins with an initial guess of the fit and iteratively updates its fit as more data (in this case, the intensity of HXR emission at a given time) is fed to the algorithm. Unlike traditional regression methods, which learn the exact values for each function parameter, GP regression models estimate a probability distribution over all possible fits.
\\
\\
A common issue with traditional methods for smoothing data in search of QPPs is the choice of window size. This choice is typically ambiguous and can lead to erroneous results if a window size is not selected manually for individual flares \citep{broomhall}. Since GP regression is based on Bayesian statistics, the model parameters can be chosen systematically by optimising a loss function, which estimates how well the model fits to the data.  Also, GP regression has the added benefit of giving an uncertainty estimate on the fitted smooth curve and, unlike most methods involving smoothing windows, GPs can be applied to unevenly sampled data. This is particularly important for STIX data, as it is binned onboard in a dynamic way to optimise the amount of down-linked data \citep{Krucker_2020}. Of course, time series can be interpolated and re-sampled at even cadence, but this introduces additional errors; thus, it is preferable to have a method that does not require such pre-processing. 

\subsubsection{GP kernel \& hyperparameter optimisation}
A GP is a collection of random variables such that any finite collection of those random variables has joint Gaussian distributions \citep{Rasmussen2004}. A GP can be entirely described by its mean and covariance function \citep{Rasmussen2004}. This is similar to a Gaussian distribution, which is defined by a mean vector and covariance matrix; however, in a GP the distribution is taken over functions \citep{Rasmussen2004}. In other words, the final best fit obtained using GP regression is given by the maximum likelihood of a multidimensional probability distribution over all possible fits. This distribution is specified by a covariance matrix (also known as a kernel). 
\\
\\
In this work, a simple radial basis function (RBF) kernel was chosen because an RBF kernel is a common choice for smooth time series data. No inference is made based on the kernel parameters themselves; the GPs are solely used as a smoothing technique. \citet{gp_for_qpps} have discussed assigning a physical meaning to kernel parameter choice in the context of QPPs. For that case, it would be important to consider various types of kernels. However, for our purpose, a simple choice of kernel was deemed sufficient. 
\\
\\
The kernel chosen for this work has two components:  a constant kernel component and a radial basis function kernel (squared exponential kernel) component. It takes the following form:
\begin{equation} \label{kernel}
     k(x_i,x_j) = Ae^{\frac{-d(x_i,x_j)^2}{2l^2}} 
,\end{equation}

where A is a constant value, $d(x_i,x_j)$ is the Euclidean distance between two feature vectors $x_i, x_j$, and $l$ is the length scale. 
\\
\\
For this kernel choice, there are two hyperparameters to optimise, $A$ and $l$; here, $l$ roughly describes the length of the (time) scale over which data points are related. For example, for a flare profile with subsecond variations, the length scale parameter is expected to be a lot smaller than for a profile with longer (on the order of seconds) variations. Then, $A$ is the amplitude of the kernel. In addition, there is a parameter $\alpha$, which is added to the diagonal of the kernel matrix to help with fitting and can be considered as Gaussian measurement noise on the training data. 
\\
\\
The choice of kernel and its parameters can be optimised by performing an exhaustive search over a wide range of values and computing the performance of the model using cross validation. This is known as hyperparameter optimisation. In cross-validation, the data is split between training and test data. The training data is used to fit model parameters and the test data is used to assess model performance. In this work the split between training and test data is 80:20.
In other words, the model is fit using the training data, and the "goodness of fit" is assessed by comparing the model prediction with the known measured value from the test set. The error on the fit is computed based on a chosen loss function -- in this case,this is the mean squared error (MSE). An average MSE value on all test data is used to assess overall model performance. Once the model hyperparameters have been optimised, a smooth time profile is obtained (as shown in Fig. \ref{fig:method_overview_plot}C).

\subsection{Fitting a linear combination of Gaussians to the HXR profile} \label{subsection:gaussian_fitting}
In many flare cases, the shape of each HXR pulsation can be aptly fit with a Gaussian function, which gives a symmetric rise and fall. The use of a simple functional form such as a Gaussian is preferable because we can easily derive insightful properties from the mean and standard deviation of each individual peak, including the waiting time between peaks, amplitude, and full width at half maximum (FWHM). To a large degree, the form of a Gaussian accurately models the sudden impulse of non-thermal electrons reaching the chromosphere and interacting to produce non-thermal bremsstrahlung emission. Of course, there are cases in which this form may not be a suitable choice; for instance, when there is clear particle trapping in the coronal loop, leading to asymmetric HXR emission profiles and a longer decay time. However, to a large extent, the choice of this functional form describes the observed HXR emission extremely well. Other reasonable choices would include a triangular pulse or a double exponential (a symmetric exponential rise and decay). In order to account for particle trapping, it could also be useful to consider an asymmetric function, such as one with an exponential rise but with a slower decay rate.
\\
\\
In this step, the smoothed signal is fit by a linear combination of Gaussian functions. The gradient of the time profile is computed and the locations where the gradient is zero are identified as peaks and troughs (see Fig. \ref{fig:method_overview_plot}D). Each pair of local maxima and minima are considered as a single Gaussian contribution. From these values, initial parameters for the fitting routine are derived. The time of the peak value is considered to be the mean of the Gaussian and the time between peak and trough is taken as a rough estimate of the FWHM. The height of the curve at the peak time is taken as an initial estimate of amplitude. Scipy's Curve Fit \citep{2020SciPy-NMeth} routine is used with the derived initial values as input and with upper and lower bounds on the possible parameter space to fit the time profile. The range of reasonable parameter values varies with each flare. A large GOES class flare will have a smaller possible range in peak height for fitting, as the counting statistics are better. 
In general, the range used is quite large to allow for the curve fitting routine to work effectively, yet it excludes wide Gaussian fits. Without putting a restriction on the range of peak widths, the curve fitting routine may return Gaussian contributions with large FWHM. This is demonstrated in Fig. \ref{fig:bounds_no_bounds}.

\section{Results}
In Sect. \ref{sec:details_04_05_22}, we present the detailed results of each step of the method for the SOL2022-05-04 flare, for the purposes of demonstration. The process has been tested on four M- and X-class solar flares observed with STIX since September 2021. In Sect. \ref{sec:flare_overview}, we give an overview of all flares analysed and the results obtained. Finally, Sect. \ref{sec:line_fit} presents an analysis of QPP identification and characterisation for each event.

\subsection{Method results detailed for the SOL2022-05-04 event}\label{sec:details_04_05_22}
\subsubsection{Gaussian process regression}

GP regression was applied to the SOL2022-05-04 flare of M1.2 GOES class. A grid search was performed over the hyper-parameter space shown in Table \ref{tab:hyperparameter_grid}. For this flare, the optimal solution was found to be $A=1.8$, $\alpha = 0.007,$ and $l=0.08$. The smoothed GP prediction for the optimised hyper-parameters is shown in Fig. \ref{fig:method_overview_plot}C with a 95\% confidence interval.

\subsubsection{Gaussian fitting}
The peaks and troughs of the smoothed SOL2022-05-04 time profile are identified (as shown in Fig. \ref{fig:method_overview_plot}D) and used as input into the fitting routine. The fitting routine fits a linear combination of individual Gaussians to the smoothed time profile. Constraints are given on the possible range of Gaussian characteristics, and for the case of SOL2022-05-04, the bounds applied are those shown in Table \ref{tab:fitting_bounds}. The resulting fit is shown in Fig. \ref{fig:method_overview_plot}E. Overall, a good fit to the data is obtained. However, at $t \approx 60$s, several peaks are not well fit. This is because the smoothing step (see Fig. \ref{fig:method_overview_plot}C) has removed some of these peaks from the time series as they are very short-lived. One of the short peaks remains in the smooth profile, although it has a smaller amplitude and, thus, it is also not very well fit in the final Gaussian decomposition. This demonstrates an important drawback of the smoothing step.
\begin{center}
\begin{table}
    \centering
    \begin{tabular}{|c|c|c|c|c|}
        \hline
         \textbf{Hyperparameter} & \textbf{Min} & \textbf{Max} & \textbf{Step} & \textbf{Solution} \\
         \hline
         $A$ & 1 & 5 & 0.1 & 1.8\\
         \hline
         $l$ & 0.01 & 0.1 & 0.01 & 0.08\\
         \hline
         $\alpha$ & 0.001 & 0.015 & 0.001 & 0.007\\
         \hline
    \end{tabular}
    \caption{Model hyperparameter grid search domain for the SOL2022-05-04 flare.}
    \label{tab:hyperparameter_grid}
\end{table}
\end{center}

\begin{center}
\begin{table}
    \centering
    \begin{tabular}{|c|c|}
        \hline
         \textbf{Parameter} & \textbf{Range}  \\
         \hline
         Mean (s) & $\pm{10}$  \\
         \hline
         Amplitude (Counts$s^{-1}$) & $\pm{200}$ \\
         \hline
         FWHM (s) & $\pm20$ \\
         \hline
    \end{tabular}
    \caption{Bounds applied to the fitting parameters for SOL2022-05-04 flare. The range used is quite large to allow for a reasonably large parameter space to be searched, but while excluding non-physical results.}
    \label{tab:fitting_bounds}
\end{table}
\end{center}

\subsection{Flare overview}\label{sec:flare_overview}
\subsubsection{SOL2021-09-23 M1.9 GOES class flare}
The SOL2021-09-23 flare was an early impulsive M1.9 GOES class flare observed by STIX when Solar Orbiter was at 0.60 AU from the Sun and at an angle of $\sim33^{\circ}$ to the Sun-Earth line. A case study of this flare by Stiefel et al., 2023 (accepted) reveals four non-thermal HXR sources observed by STIX, with two inner sources from the traditional flare loop and outer footpoints which may be associated with the early onset of a filament eruption; however, the evidence on the latter is inconclusive. A Gaussian decomposition for this flare was performed and the resulting fit (shown in Fig. \ref{fig:230921_gaussian_decomp}) has been assessed by calculating the standard deviation of the normalised residual. The normalised residual is calculated as the difference between the fit and the observed STIX time profile normalised by the error on the measured STIX time profile. The error on the STIX observed time profile is a combination of the compression error and counting statistics error. The fit gives $\sigma_R = 1.83$. The residuals are quite high, particularly at the beginning of the impulsive phase because there are fast-time variation on timescales smoothed out by the Gaussian process regression. However, the fit improves in the later phase. 

\subsubsection{SOL2021-10-09 M1.6 GOES class flare}
The SOL2021-10-09 flare was an M1.6 GOES class flare observed when Solar Orbiter was at a distance of 0.68 AU from the Sun with the spacecraft-Sun-Earth angle of $\sim 15^{\circ}$. The Gaussian decomposition method gives a fit with standard deviation of the residual $\sigma_R=0.62$, shown in Fig. \ref{fig:091021_gaussian_decomp}. In this case, the measured data is noisier than other example flares because the flare is smaller. 
As a result, it is easier for the fit to appear to perform well on global structures, but the variance among the fits is high.

\subsubsection{SOL2022-03-30 X1.4 GOES class flare}
\begin{figure*}
    \centering
    \includegraphics[width=\textwidth]{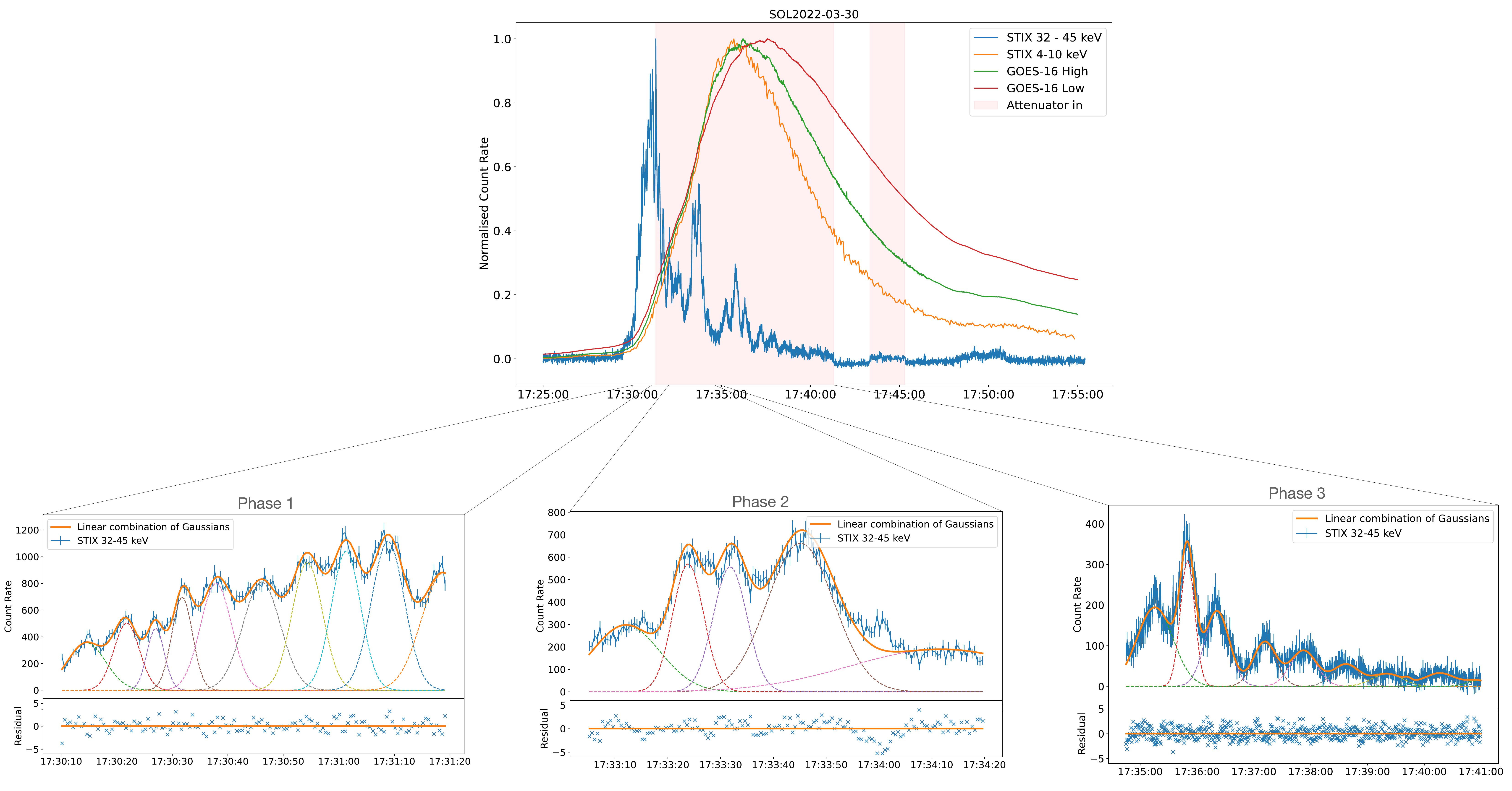}
    \caption{Overview plot of the SOL2022-03-30 flare and the Gaussian decomposition of the 32-45 keV HXR profile for three different phases: phases 1, 2, and 3 from left to right. The standard deviation of the residual of each phase is $\sigma_{R,1} = 1.25$ , $\sigma_{R,2} = 1.58$ \& $\sigma_{R,3} = 1.11$.}
    \label{fig:300322_gaussian_decomp}
\end{figure*}

The X1.4 GOES class flare observed on 30-03-2022 was the first X class flare to be observed by STIX during perihelion. This observation was taken at a distance of 0.33 AU from the Sun. Furthermore, the angle between Solar Orbiter and the Sun-Earth line at this time was $\sim 95^{\circ}$. Due to the large flux incident on the instrument during this flare, the aluminium alloy attenuator was automatically inserted when a certain trigger threshold was reached. The periods when the attenuator was inserted are shown in Fig. \ref{fig:300322_gaussian_decomp}. The attenuator mainly blocks low energy X-ray photons and allows high energy photons to be transmitted. The energy dependent response of the attenuator can be found in \citet{Krucker_2020}. As the counts incident on the detector increase, the live-time decreases and vice versa when the attenuator is inserted. Therefore, to correctly show the time profile of such a flare, we should carefully account for changes in detector live-time as well as the energy and spectrum dependent attenuator transmission. This correction is non-trivial as it requires detailed knowledge of the intricacies of such an instrument. However, for the background detector (BKG), the pixels with large apertures are turned off when flux is high. This means that the attenuator does not cover this detector when inserted. As such, there is no effect exerted by the attenuator motion on the 4-10 keV background detector time profile shown. The background detector was used for the thermal 4-10 keV profile shown in Fig. \ref{fig:300322_gaussian_decomp}, as it requires no correction for the attenuator motion.
\\
\\
This flare is an early impulsive flare which shows time-varying structures. In particular, there are variations on the timescale of several seconds. Later in the flare, during the thermal peak, there are further variations with significantly different fluctuation timescales ($\sim 14$s) and, later on, even longer periods of $\sim 35$s. The pulsation timescales are changing significantly over the course of the flare and so, it is challenging to smooth the time profile using our simple choice of kernel, since the length-scale parameter has no time dependence. To account for this, the time profile was split into three different sections: the early phase (phase 1), middle phase (phase 2), and late phase (phase 3). Figure \ref{fig:300322_gaussian_decomp} shows the Gaussian decomposition for each phase with the standard deviation of the residuals being $\sigma_{R,1} = 1.25$ , $\sigma_{R,2} = 1.58$ \& $\sigma_{R,3} = 1.11$ for phases 1, 2, and 3, respectively. In particular, phases 1 and 3 appear to be well fitted with a linear combination of Gaussians, but the residuals for phase 2 show sinusoidal variation, indicating the data are not well fit by the model.

\subsubsection{SOL2022-05-04 M1.2 GOES class flare}
\begin{figure}
    \centering
    \includegraphics[width=0.5\textwidth]{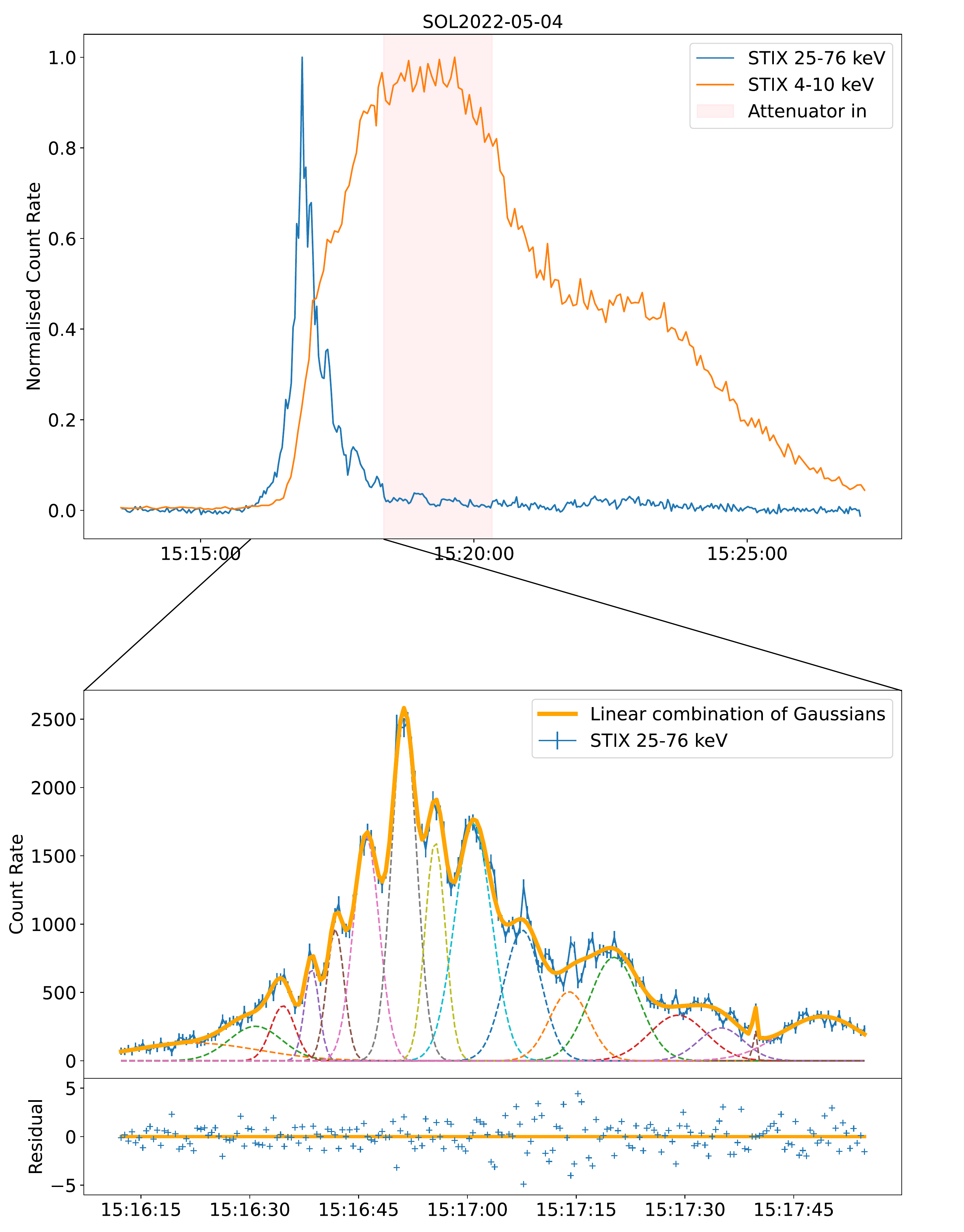}
    \caption{Overview plot of the SOL2022-05-04 flare and the Gaussian decomposition of the impulsive phase 25-76 keV profile. The standard deviation of the residual for the fit is $\sigma_R = 1.39$.}
    \label{fig:040522_gaussian_decomp}
\end{figure}

Finally, we present the M1.2 GOES class flare on May 4th 2022. At this time, Solar Orbiter was observing the far side of the Sun from the Earth, namely, the spacecraft-Sun-Earth angle was $\sim 163^{\circ}$ and the distance to the Sun was 0.73 AU. As such, the flare was not observed by Earth based observatories. The GOES class of a flare which was not observed from Earth is estimated using a model that fits STIX 4-10 keV counts to the GOES flux\footnote{\url{https://datacenter.stix.i4ds.net/wiki/index.php?title=GOES_Flux_vs_STIX_counts}}. Similarly to the SOL2022-03-30 flare, the attenuator motion is shown in Fig. \ref{fig:040522_gaussian_decomp} and the background detector was used for the thermal 4-10 keV profile. The standard deviation of the residual $\sigma_R = 1.39$. Overall, the data is well fit with a linear combination of Gaussians. The fastest fluctuations in time are not well fitted (e.g. Fig. \ref{fig:method_overview_plot}E at $ t \approx 60$s) because the time profile output by the GP regression smooths out some very short variations. Therefore, the Gaussian fitting procedure doesn't fit these shorter peaks. This is a limitation of the method that is discussed further in Sect. \ref{section:drawbacks}. 

\begin{figure}
    \centering
    \includegraphics[width=0.49\textwidth]{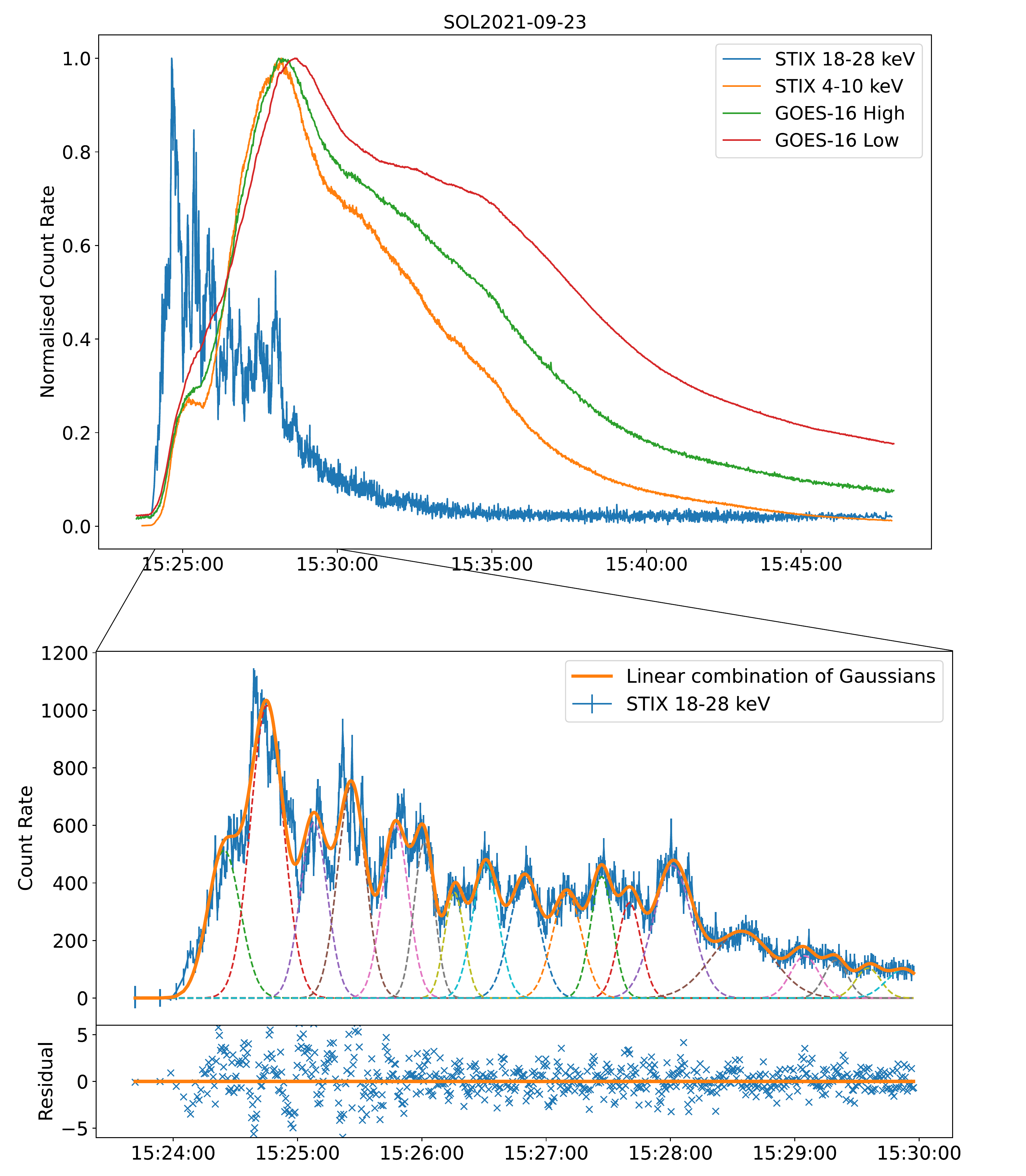}
    \caption{Overview of the SOL2021-09-23 flare. The 18-28 keV HXR non-thermal Gaussian decomposition is shown with a standard deviation of the residual $\sigma_R = 1.83$. The first few peaks have substructure on timescales smoothed out in the GP regression output. As a result, variations on these timescales are not well
  fit.}
    \label{fig:230921_gaussian_decomp}
\end{figure}
\begin{figure}
    \centering
    \includegraphics[width=0.49\textwidth]{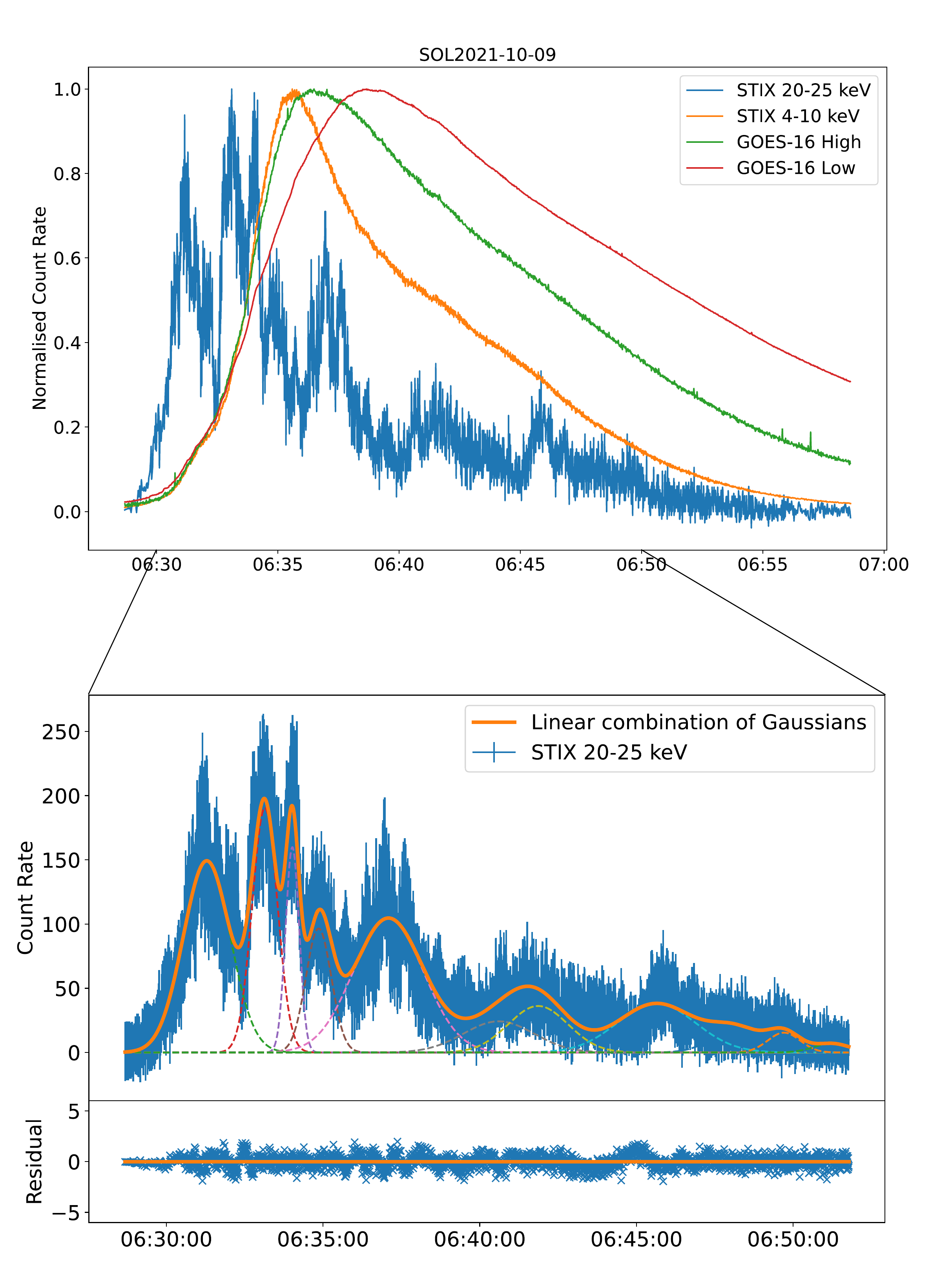}
    \caption{Overview of the SOL2021-10-09 flare and the Gaussian decomposition of its 20-25 keV HXR non-thermal time profile. The fit gives a standard deviation of the residual $\sigma_R = 0.62$. This is a strong fit. The noise level on the measured values of the counts is higher for this flare since it is smaller. This makes it easier to fit and thus the residuals are smaller than those of a large flare with higher signal-to-noise ratio.}
    \label{fig:091021_gaussian_decomp}
\end{figure}

\subsection{QPP identification and characterisation}\label{sec:line_fit}

\begin{figure*}
    \centering
    \includegraphics[width=\textwidth]{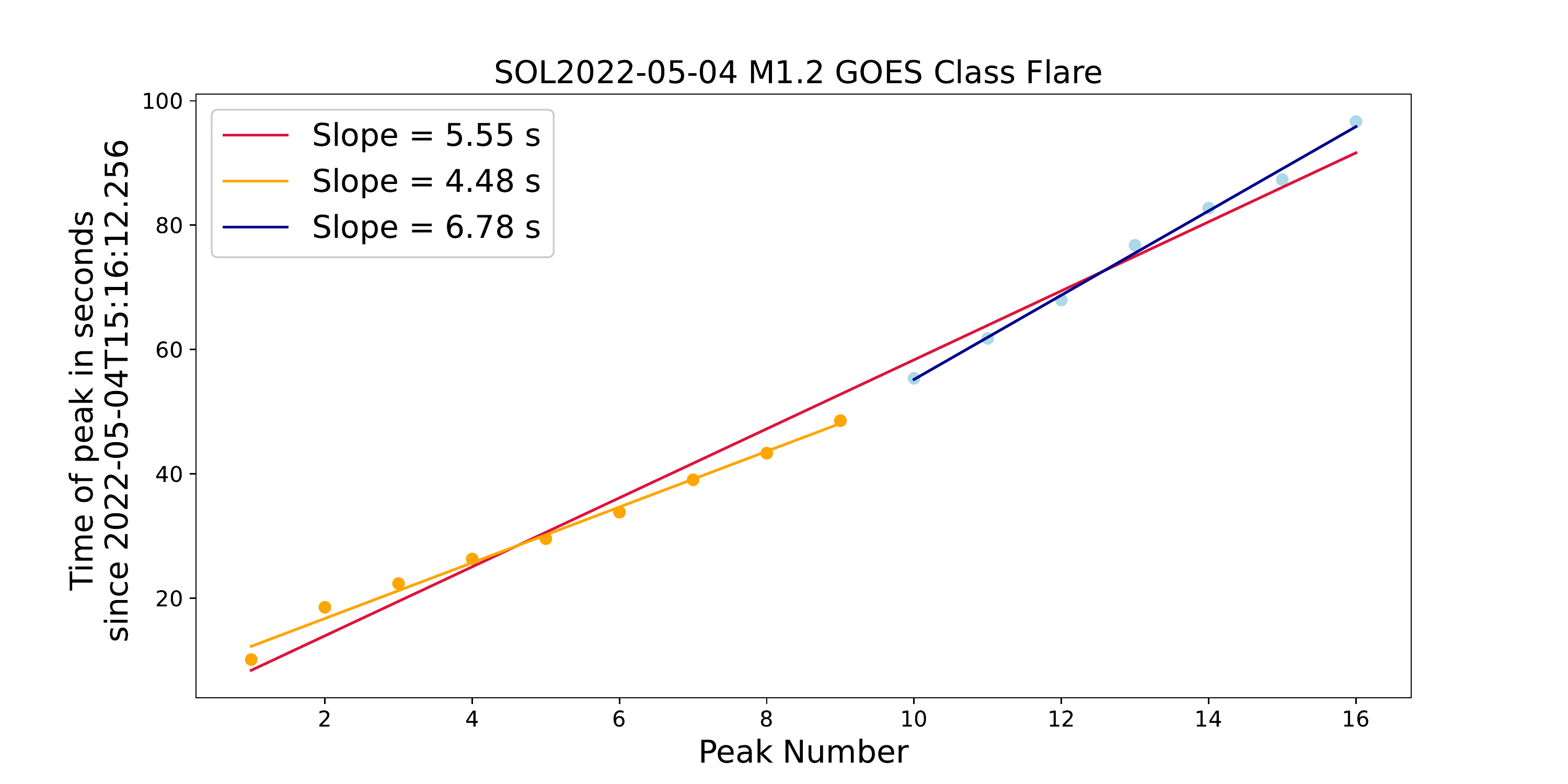}
    \caption{Mean time of each Gaussian component against peak number. The slope of the line fit gives an estimate of periodicity in the signal. A stronger fit with multiple lines would indicate that there is non-stationarity in the signal.}
    \label{fig:3linefit}
\end{figure*}

An estimate of signal (quasi-) periodicity can be derived from individual Gaussian components. Figure \ref{fig:3linefit} shows the time of each Gaussian component against the peak number for the SOL2022-05-04 flare. The slope of the line of best fit gives an estimate for the period, $\sim 5.55$s. This shows that STIX can detect variations on short timescales, which is something that was not feasible with its predecessor RHESSI. This result is then compared with the AFINO analysis method \citep{inglis2017}. AFINO fits the Fourier power spectrum of a flare signal with four different models: 1) a power law + constant; 2) power law + constant + Gaussian
    A broken power law + constant; 3) power law + 2 Gaussians. 
    Should a QPP component be present in the signal, the second or fourth model would be favoured, that is, the Gaussian peak fits the enhanced power due to the QPP that is present. The goodness of fit is estimated from a Bayesian information criterion (BIC), given by:  
\begin{equation}
    BIC = -2 ln(L) + k ln(n),
\end{equation} 
where $L$ is the maximum likelihood, $k$ is the number of free parameters, and $n$ is the number of data points in the Fourier power spectrum. Thus, the BIC score penalises over-fitting. A large negative value for a given model indicates a strong fit to the data. One model is said to be strongly preferred over another if $\left|{\Delta{BIC}}\right| > 10$. In the case of the SOL2022-05-04 flare, the BIC score for model 1 versus model 2 is $\left|{\Delta{BIC}}_{12}\right| = 4.3$, whereas model 2 gives a larger negative BIC score. Thus, the AFINO method gives a slight preference for the QPP model with single period, $P= 4.96^{+0.66}_{-0.54}$s, over a simple power law model. This is largely consistent with the period obtained from the Gaussian decomposition method, which is slightly higher since smoothing the time profile suppresses the fastest time variations. We notice that when assessed based on the standard AFINO criteria ($\left|{\Delta{BIC}}\right| > 10$), this flare would not be marked as a QPP detection, although there is clearly enhanced power at $f_0 = 0.202 \pm 0.024 $ Hz ($P = 4.96^{+0.66}_{-0.54}$s). From the line fitting method, there is the added benefit of obtaining the exact timing information of each pulsation, which can be used to spatially resolve each peak. 
\\
\\
By fitting two lines, a test can be performed to check for non-stationarity (changes in quasi-periodicity over time) in the flare signal. The derived slopes are $4.48$s and $6.78$s for the early and late stages, respectively. However, the Pearson correlation co-efficients obtained indicate that a single line fit or single periodicity in the signal gives a better fit to the data. This is, of course, biased by the number of data points in each fit. Further analysis was done where the signal was split into two time ranges, corresponding to the time ranges of the two line fits. AFINO analysis was performed on these two time ranges. In this case, the QPP model was only favoured in the first time range. Therefore, the AFINO method favours a single QPP model over two periods in this particular case. This agrees with the apparent preference for a single periodicity given by the line fit.
\\
\\
It has been demonstrated that this new method is capable of accurately detecting and characterising fast-time-varying structures and QPPs in the non-thermal HXR emission from flares. The results from this method are consistent with Fourier based analysis methods such as AFINO and in addition, allow for the extraction of important information, for instance, the time between peaks and pulsation duration. Further, this method can be used to detect frequency drifts and non-stationarity in time series.

\begin{figure}
    \centering
    \includegraphics[width=0.5\textwidth]{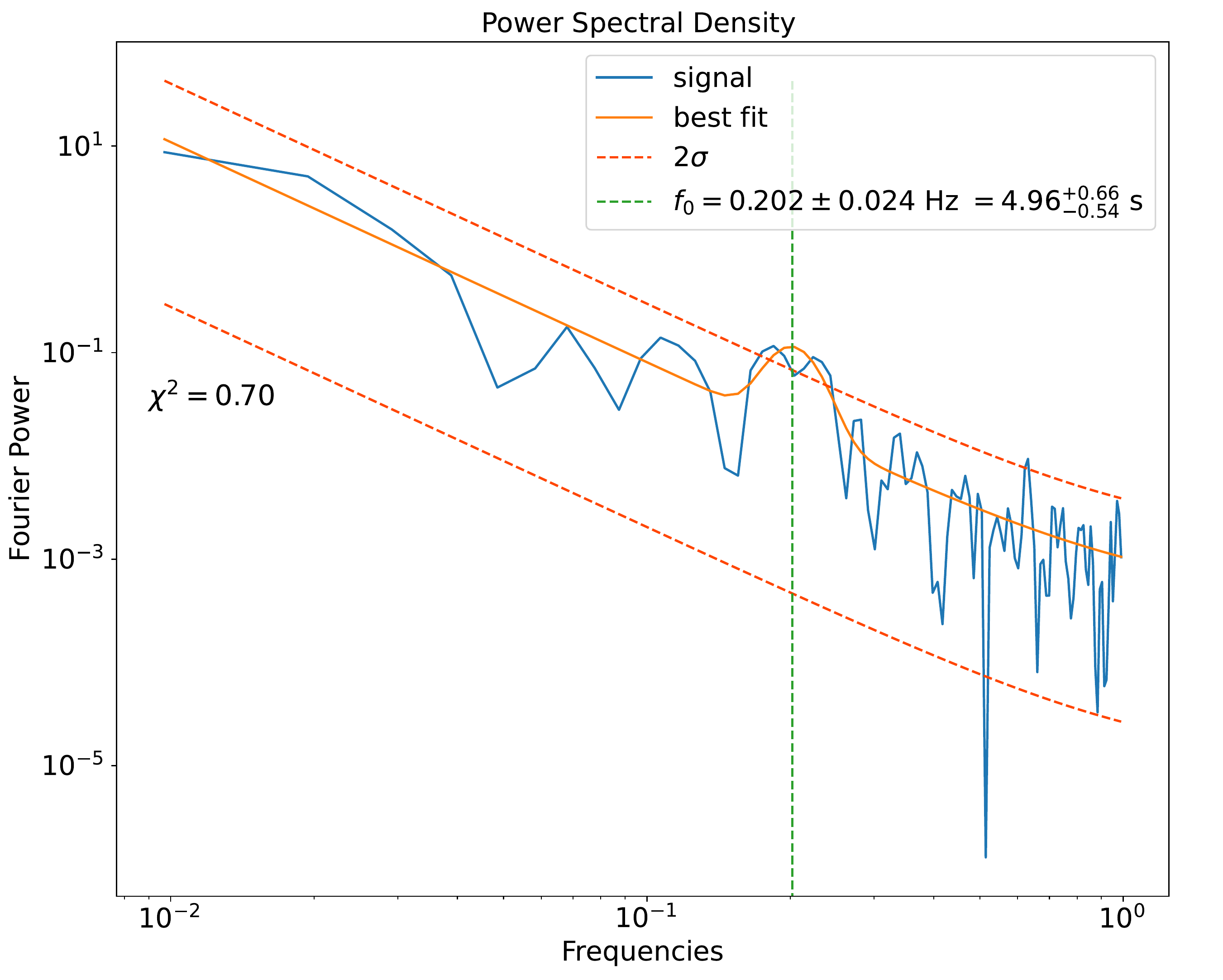}
    \caption{Power spectral density of signal shown in Fig.\ref{fig:method_overview_plot}A. The AFINO QPP model best fit is shown. AFINO analysis gives a period of $\sim4.96^{+0.66}_{-0.54}$s, with a moderate preference for the QPP model with $\left|{\Delta{BIC}}\right| = 4.3$}
    \label{fig:afino}
\end{figure}

\begin{figure*}[htp]
  \centering
  \subfigure{\includegraphics[width=0.49\textwidth]{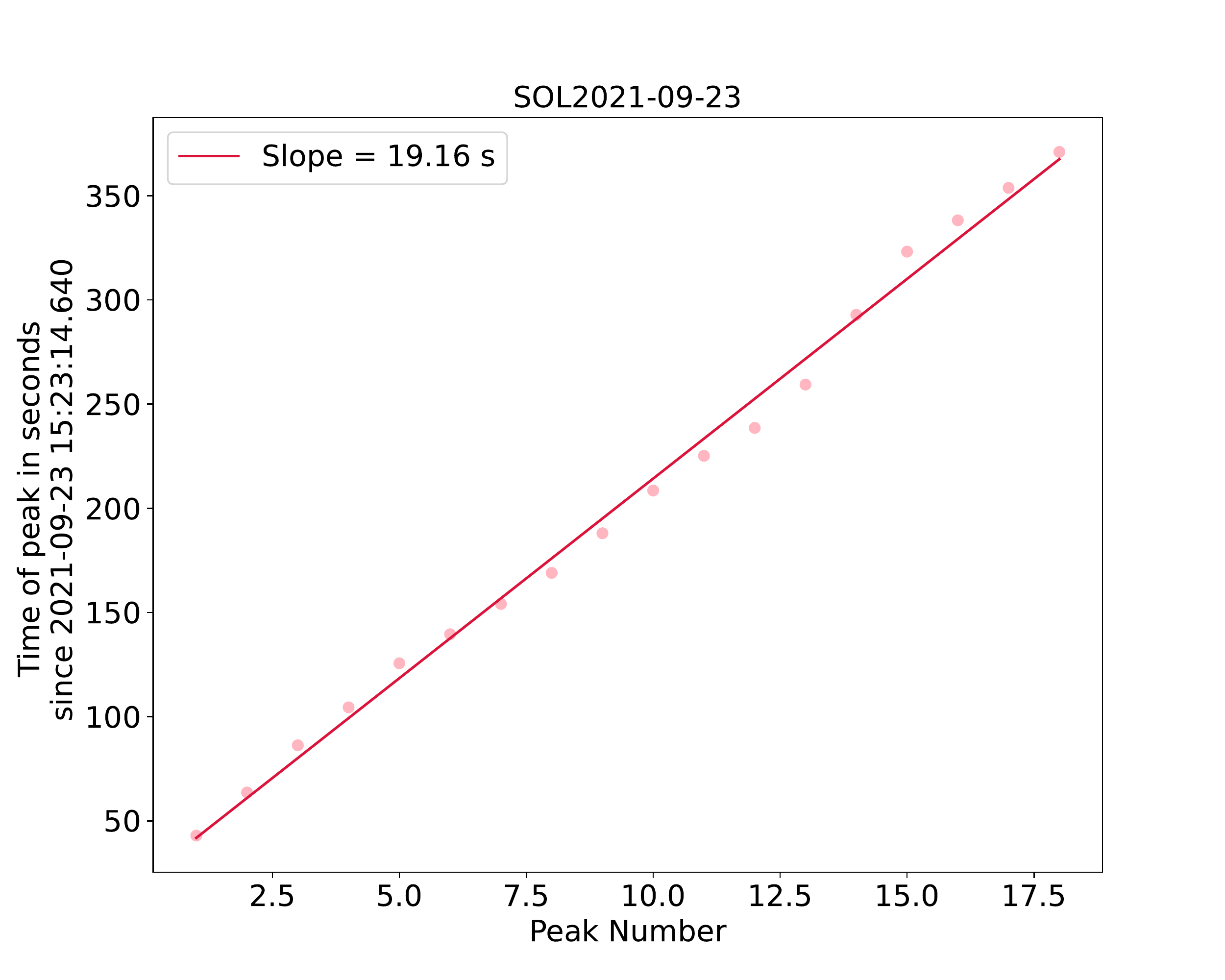}}\quad
  \subfigure{\includegraphics[width=0.49\textwidth]{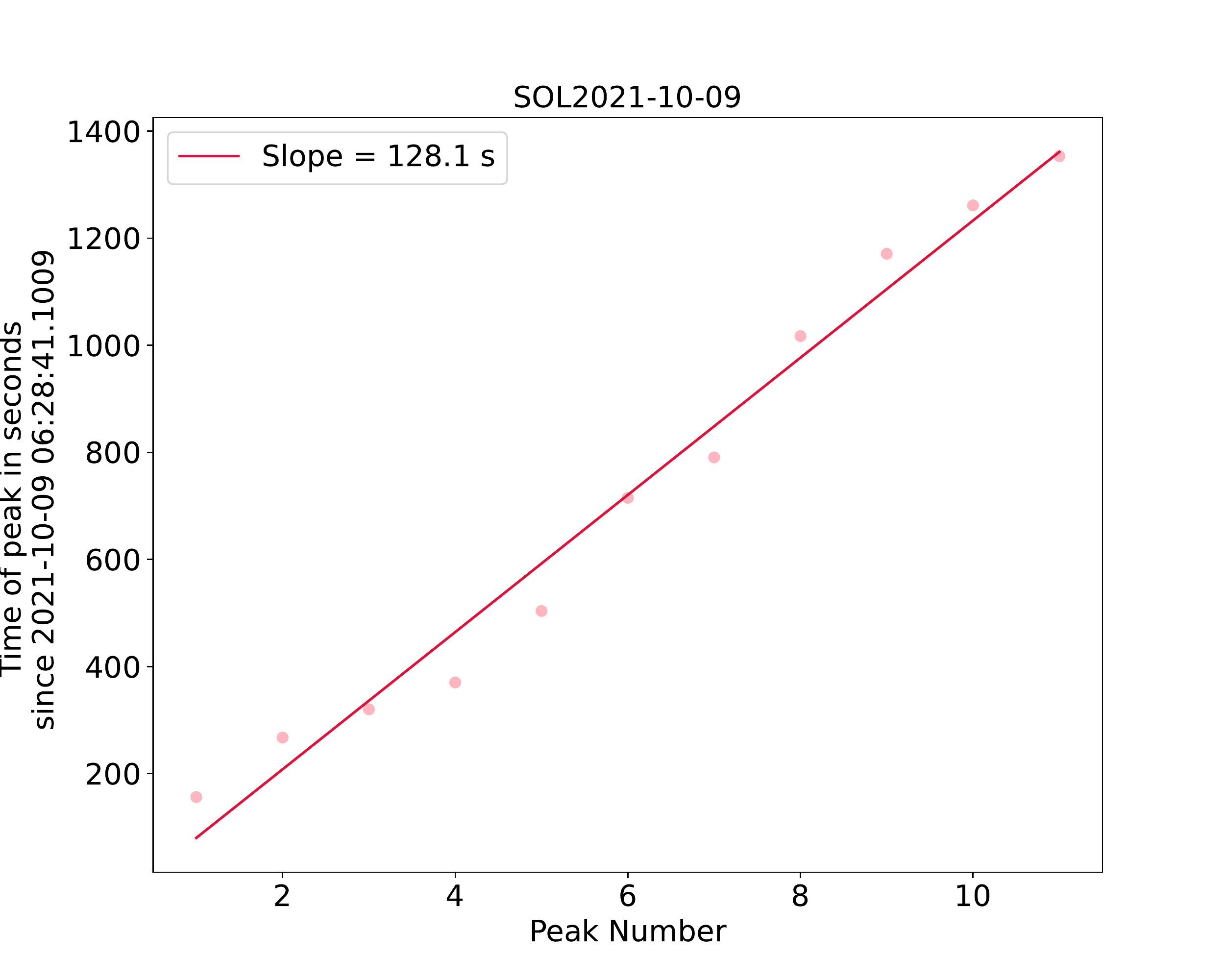}}

  \caption{Time of each peak from the Gaussian decomposition against peak number for the SOL2021-09-23 \& SOL2021-10-09 flare shown in Figs. \ref{fig:230921_gaussian_decomp} and \ref{fig:091021_gaussian_decomp}, respectively. The fit gives an estimate of periodicity in the SOL2021-09-23 non-thermal HXR flare signal of $\sim 19$s. For the SOL2021-10-09 flare, a periodicity of $\sim 128$s is derived.}
\end{figure*}

\begin{figure*}[htp]
  \centering
  \subfigure{\includegraphics[width=0.32\textwidth]{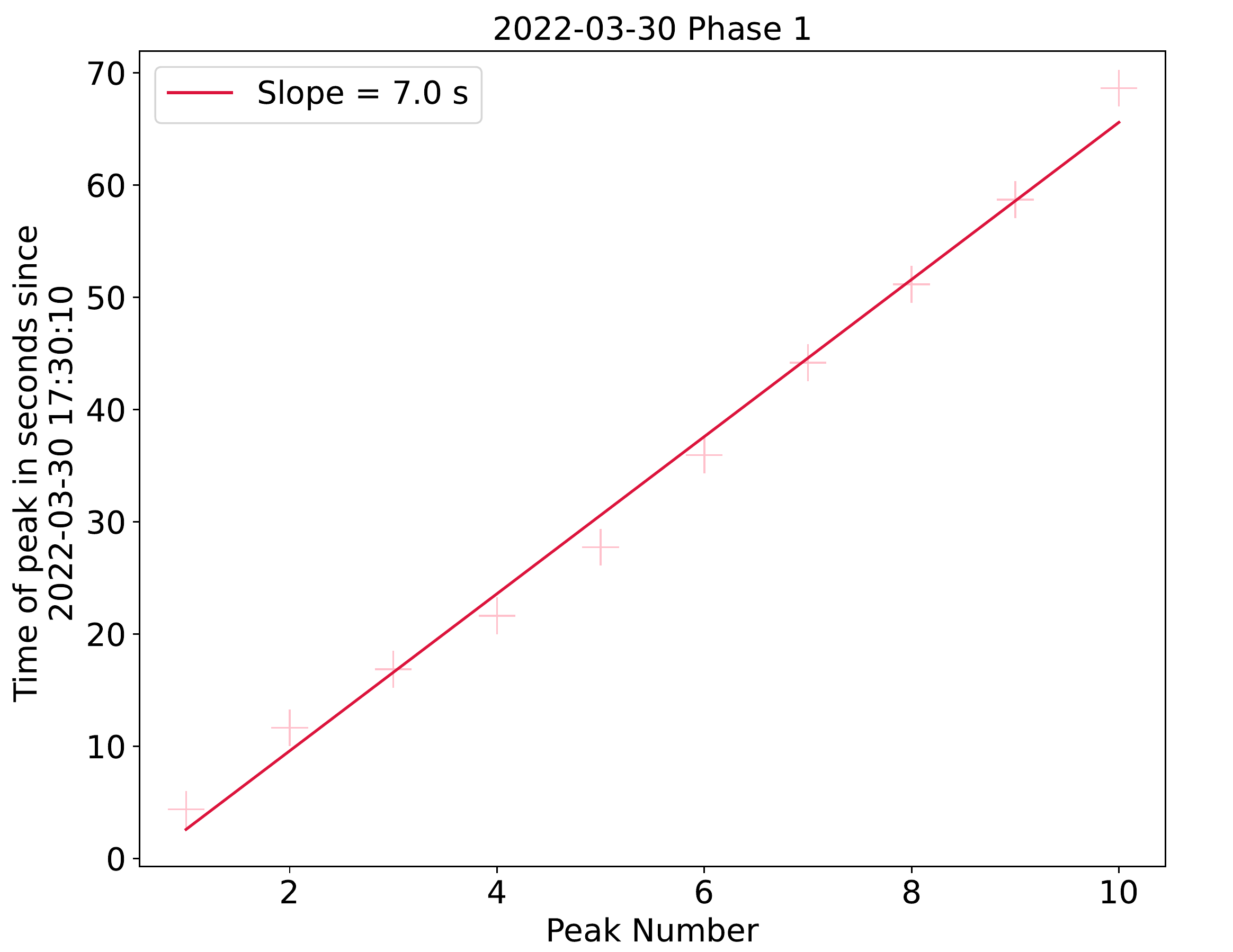}}\quad
  \subfigure{\includegraphics[width=0.32\textwidth]{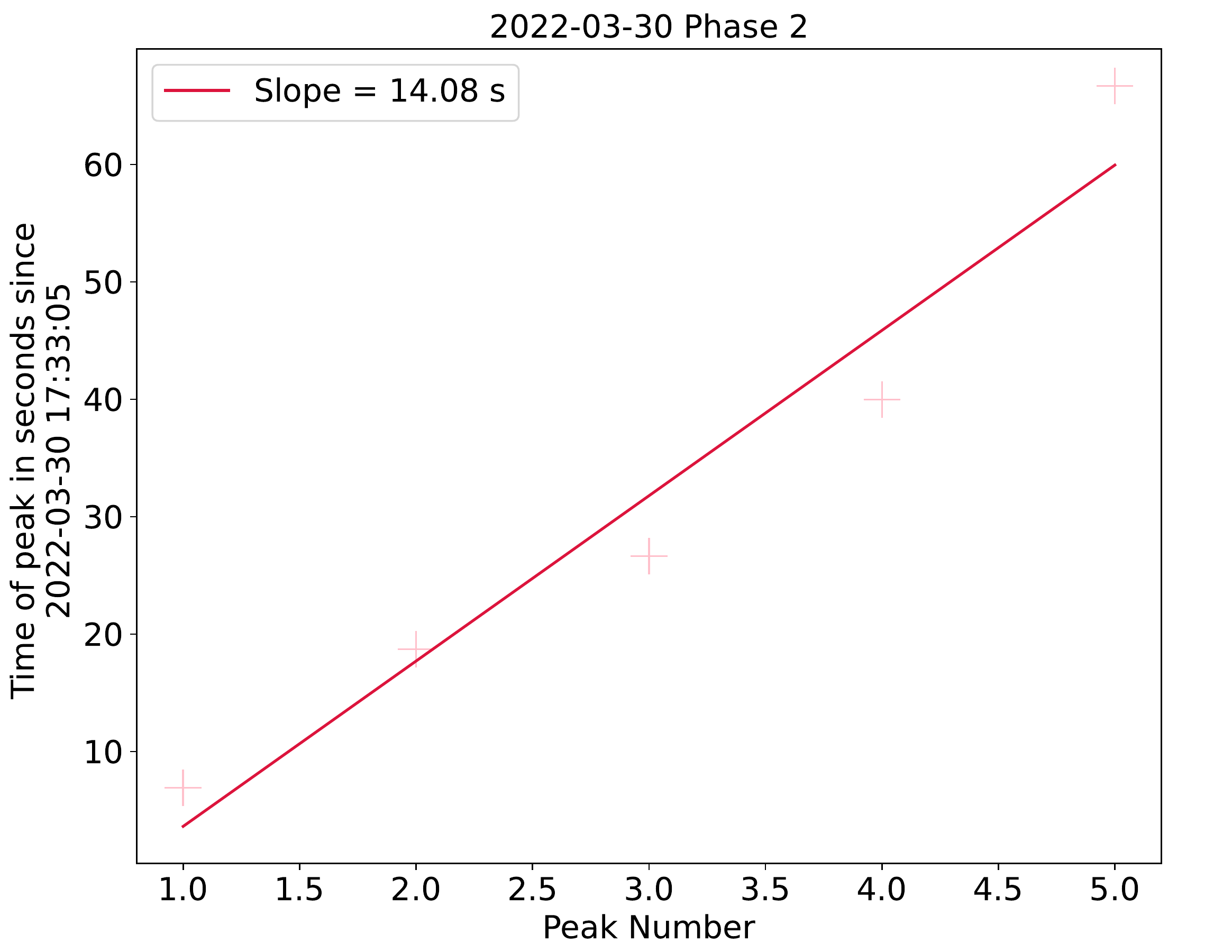}}\quad
    \subfigure{\includegraphics[width=0.32\textwidth]{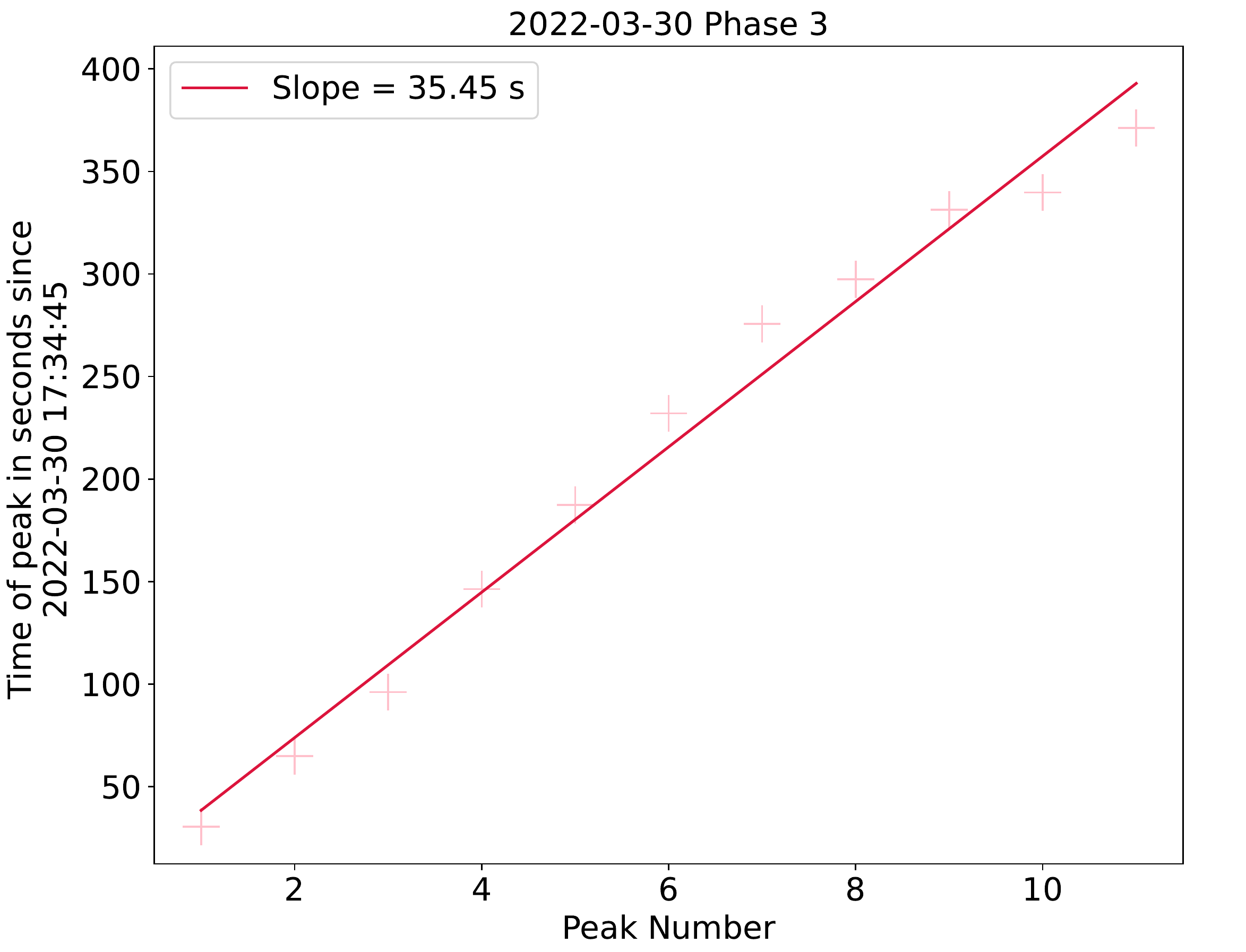}}
  \caption{Line fits to the peak time over peak number for three phases of the SOL2022-03-30 flare shown in Fig. \ref{fig:300322_gaussian_decomp}. Since the pulsation frequency is increasing significantly in time, each phase was analysed separately. Phase 1 is very well fitted by a line indicating that the timing of peaks is quasi-periodic, phases 2 
    and 3 are not as well fit by a line. This agrees with the error analysis presented in Table \ref{tab:av_period}, where phases 2 and 3 have high variance. Between phases 1 and 3, the periodicity has increased by a factor of $\sim5$. This indicates that the driver of these pulsations could be different in the later stages of the flare, compared to the initial impulsive phase.}
    \label{fig:300322_3linefits}
\end{figure*}

\subsubsection{Errors on QPP analysis}
The Gaussian decomposition fits are non-unique and multiple solutions can be derived for a given time series. To determine the effect of this, error analysis was performed for determining the QPP period. For this, an array of errors was added to each STIX time series, which is some random integer multiple of the calculated measurement error (that includes the uncertainty introduced due to compression and counting statistics). The random integer is sampled from a normal white noise distribution with zero mean and unit standard deviation. The entire fitting procedure is then performed 50 times, with different initial error arrays added to the time profile. This gives many unique Gaussian decomposition fits for a given flare profile. As previously, the periodicity of the signal is estimated from a single line fit and an average period for each flare is obtained, with a standard deviation over 50 iterations. 
The mean periods and standard deviation are as shown in Table \ref{tab:av_period}. An estimation of the range of periods that reasonably describe the variations observed is also given. 

\begin{table}
    \centering
    \begin{tabular}{|c|c|c|}
        \hline
        \textbf{Flare} &  \textbf{Mean period (s)}&  \textbf{Range (s)}\\
        \hline

        SOL2021-09-23 & $18.5\pm  0.7$ & $17.5-19.9$ \\
        \hline
        SOL2021-10-09 & $100.9\pm 9.5$ & $82.6-129.4$ \\
        \hline
        SOL2022-03-30 P1 & $7.2\pm 0.4$ & $6.9-8.0$
        \\
        \hline
        SOL2022-03-30 P2 & $18.4\pm 3.0$ & $13.6-28.2$ \\
        \hline
        SOL2022-03-30 P3 & $35.8\pm 3.6$ 
        & $27.9-44.5$ \\
        \hline
        SOL2022-05-04 & $6.0\pm 0.4$ & $5.3-7.0$
         \\
       \hline
    \end{tabular}
    \caption{Table showing the mean and range of periods derived for each flare over 50 iterations, where random error has been added to the initial time series.}
    \label{tab:av_period}
\end{table}
\noindent
These results indicate that the period estimation for flares SOL2021-09-23, SOL2022-30-03, and  SOL2022-05-04 in Fig. \ref{fig:230921_gaussian_decomp}, \ref{fig:300322_gaussian_decomp} P1, and  \ref{fig:040522_gaussian_decomp}, respectively, are strong fits, since the estimated periods are within $1.1\sigma$ of the means obtained from the error analysis. However, the SOL2021-10-09 flare has high variance when noise is added to the time profile. This is because the HXR counts are low and as a result the measurement error is relatively high; therefore, when additional error is added, the Gaussian decomposition fits vary significantly. The fit obtained in Fig. \ref{fig:091021_gaussian_decomp} is thus unreliable, as the derived period is $2.9\sigma$ from the mean period over 50 iterations. We notice that for the SOL2022-03-30 flare, phase 1 gives a strong fit; however, phases 2 and 3 have a wide range of periods -- thus, they are not as well constrained. 

\section{Discussion and conclusions}
\subsection{First analysis of fast time-varying structures and QPP detection with STIX data}
In this work, we present the first analysis of fast-time-varying structures in the non-thermal HXR emission from flares using Solar Orbiter's STIX instrument. We have developed a new method for identifying and quantifying fast-time-varying structures in the HXR emission from flares. This method decomposes the non-thermal HXR emission from flares in the impulsive phase into a linear combination of Gaussian pulses. For a sample of four M- and X-class flares, the standard deviation of the normalised residual is $\leq 1.8$. This indicates that the model is a good fit to the data for this selection of events.
\\
\\
From these four flares, fast-time variations on timescales ranging from 4-128 seconds have been characterised. Furthermore, the first detection of solar flare QPPs from STIX observations has been made. It has been shown that QPPs with timescales down to the order of $\sim4$s can be detected with STIX, which was not possible with it predecessor RHESSI due to the spacecraft rotational period of $\sim4$s. This work demonstrates that STIX is an instrument well suited to the detection of fast-time variations in the HXR emission from solar flares in the timescale range of seconds to minutes, due to its high time resolution and relatively constant non-solar background. 

\subsection{Drawbacks} \label{section:drawbacks}
In this section, we present and discuss the drawbacks and subtleties associated with the method and how they may impact the derived results:\ 
\begin{enumerate}
    \item One important drawback of this method is that GP regression smooths time variations on timescales that are shorter than the optimal length scale obtained from hyperparameter optimisation. This means that some peaks that are very short-lasting are smoothed out and, hence, they are not fit by the Gaussian fitting routine. Importantly, the timescales that are suppressed are a function of the sampling rate, since with a higher time cadence, shorter variations will likely be more prominent and vice versa. As such, the drawback of this method is that variations that occur at the limit of the instrument's sampling cadence are smoothed out. In this case with STIX, variations on short timescales such as second or sub-second are smoothed out. While QPPs can occur on short timescales such as second or sub-second periodicities, particularly in the radio band \citep{2012ApJ...749...28T, 2018ApJ...859..154N, Carley2019NatCo..10.2276C}, this work focuses on QPPs that have timescales on the order of seconds to minutes, which are very commonly reported (see also \citet{mclaughlin2018, Hayes_2020, zimovets2021}). We have shown here that the method works well for these types of QPPs.
    \item It is well known that in the case of significant particle trapping in a flare loop, the HXR time profile becomes asymmetric. In this case, fitting Gaussian curves to the time profiles of a trapped particle population is no longer physically meaningful. We should consider fitting other functional forms with asymmetric profiles.
    \item A subtlety of this method is that although the Gaussian decomposition step is good at characterising non-stationarity in a signal, GP regression with our choice of kernel, is unable to model large frequency drifts in a flare since the length scale is not time-dependent. This gives some limitation to the method when identifying non-stationarity. This effect is particularly pertinent to the case of the SOL2022-03-30 flare, whereby in the early impulsive phase we observe fluctuations on the order of $\sim 7$s and later during the thermal peak there are fluctuations on a $\sim15$s and then a $\sim35$s timescale, as shown in Fig. \ref{fig:300322_3linefits}. One way to rectify this is to split the time series into sections and perform the smoothing on different time ranges separately, as was done for the SOL2022-03-30 flare. Another possibility is to consider a more complex kernel choice which has some time dependency. This method was also suggested by \citet{gp_for_qpps}.
    \item Another important note to consider is that the Gaussian decomposition fit is non-unique and influenced by the bounds applied to the curve fitting routine. Furthermore, the routine fits a fixed number of Gaussians based on the number of local maxima and minima. As such, several closely separated peaks in quick succession may not be fit by this method, if the gradient remains non-zero. Figure \ref{fig:bounds_no_bounds} demonstrates the effect of applying boundary conditions on the fitting parameters for the SOL2022-05-04 flare.  
\begin{figure*}[htp]
  \centering
  \subfigure[]{\includegraphics[width=0.49\textwidth]{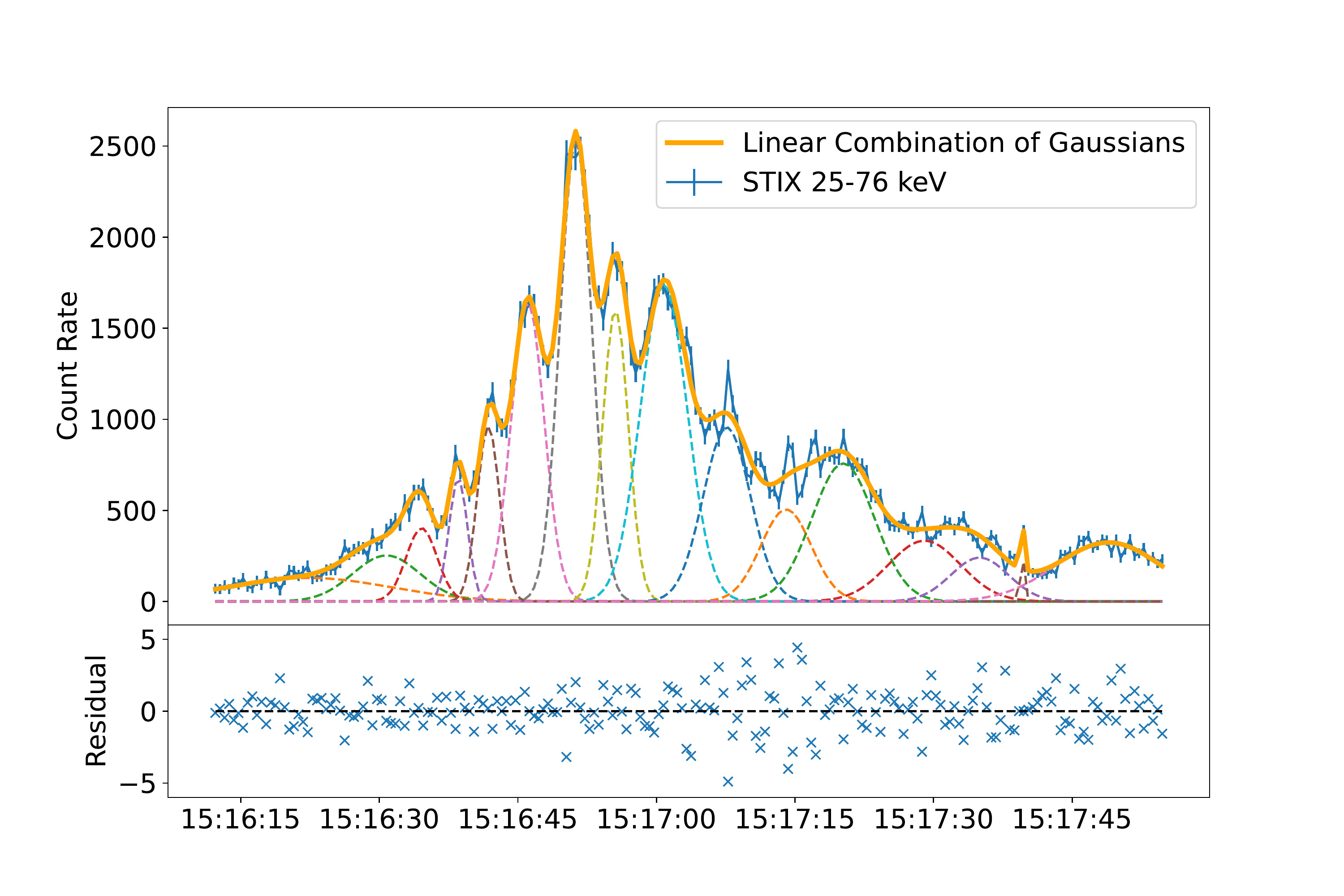}}\label{subfig:constrained}\quad
  \subfigure[]{\includegraphics[width=0.49\textwidth]{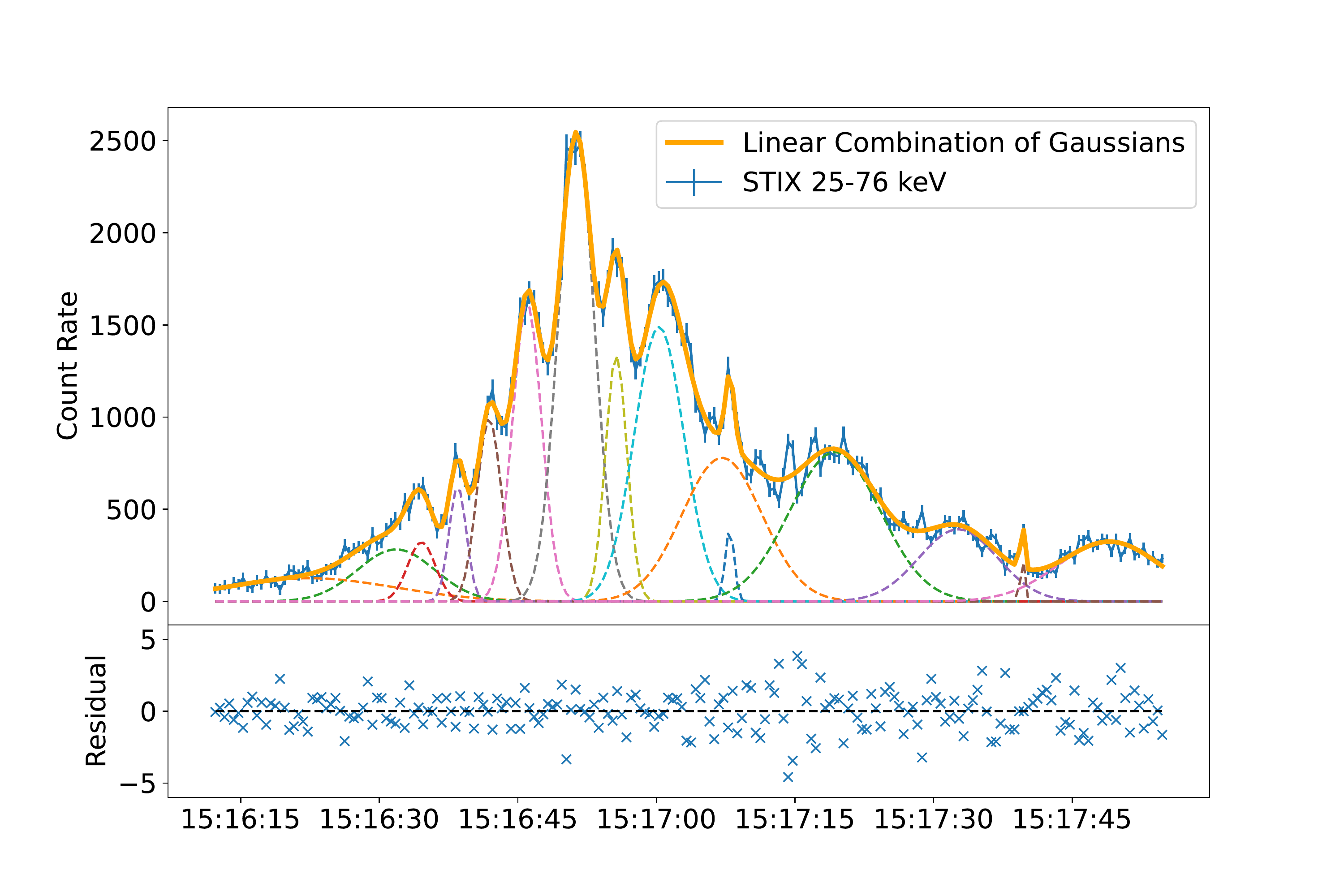}}\label{subfig:unconstrained}
  \caption{Effect of applying boundary conditions on the fit parameters of the Gaussian decomposition for the SOL2022-05-04 flare. Panel (a) shows the resulting fit when the boundary conditions shown in Table \ref{tab:fitting_bounds} are applied. Panel (b) shows the resulting fit when no boundary conditions are applied. Both cases give a good fit to the data, however, the unconstrained fit (Fig. \ref{fig:bounds_no_bounds} b) fits long period Gaussians with small amplitude peaks, for instance, at $\sim$15:17:08. This fit is interesting from a physics point of view, however in this work we wish to derive timing information regarding the global peak trend -- therefore a constrained fit is preferable.}
  \label{fig:bounds_no_bounds}
\end{figure*}
\end{enumerate}

\subsection{Applications and future work}
The potential applications of this method are wide and varied. Obtaining timing information on individual non--thermal HXR pulses in a systematic way enables a wide-scale, in-depth study of fast-time-varying structures. In particular, the timing information obtained in this method can be used to image individual pulse contributions to the HXR flare profile. With such information, we can begin to understand the spatial structure (location, morphology) of the oscillatory source, as suggested by \citet{zimovets2021}, and make a comparison with those expected for various models. For example, should sausage mode oscillations be responsible for fast-time variations in HXR profiles, we would expect the source morphology to change over time and the location to remain fixed. In contrast to this, if repeated reconnection were to be responsible for such modulations, we would expect the HXR source spatial locations to change in time since the reconnecting field lines must change. The time evolution of such structures can be investigated with STIX's imaging capabilities. Although it should be noted that imaging fast-time-varying HXR structures can be challenging with an indirect imager such as STIX due to its limited dynamic range; in particular, in cases where\ there is one source that is much brighter than the others. Furthermore, one can perform time dependent spectral analysis with STIX. This can help to identify and investigate the underlying particle population behind such oscillatory signatures. Additionally, time-dependent imaging spectroscopy can be used to localise particle populations.
\\
\\
This method adds to the set of current QPP detection techniques, with the additional ability of characterising non-QPP, fast-time-varying signatures. The quantities and relationships derived by this method are important as they feed back into modeling efforts. For example, large-scale waiting
time distributions could be used to assess whether avalanche models can accurately describe the observed fast-time variation phenomenon. 
\\
\\
Future work will focus on using the derived timing information and shape to determine over which interval HXR images should be made in order to understand the source origin. As STIX is an indirect Fourier imager, another interesting avenue of investigation will be to analyse the evolution of Fourier components (visibilities) over time. We will focus on combining the analysis with other datasets at various wavelength of emission. This method will be applied to other HXR datasets such as FERMI Gamma-Ray Space Telescope \citep{meegan2009} and RHESSI \citep{Lin2002SoPh..210....3L}, and to future instruments including ASO-S/HXI \citep{HXI_Zhang2019} and Aditya-L1/HELIOS \citep{2017aditya}. This method will also be applied to other wavelengths of emission such as pulsations observed in radio by the Expanded Owens Valley Solar Array (EOVSA) \citep{Gary_2018}, which are often correlated with those seen in HXR emission \citep{1990A&A...229..206A}.
\\
\\
In conclusion, a new method for identifying and characterising fast-time variations in the non-thermal HXR time profiles of solar flares has been developed, in which the signals are decomposed into individual Gaussian contributions. The fits obtained have a standard deviation of the normalised residual of $\leq 1.8$. The first characterisation and detection of fast-time variations in the HXR profile of solar flares with STIX has been made on timescales between 4-128s. 
\\
\\
The opportunity to study time variations in flares has greatly improved with new observations from STIX on Solar Orbiter and its observations of numerous flares that demonstrate fast-time variability over a wide range of timescales. 

\begin{acknowledgements}
           Solar Orbiter is a space mission of international collaboration between ESA and NASA, operated by ESA. The STIX instrument is an international collaboration between Switzerland, Poland, France, Czech Republic, Germany, Austria, Ireland, and Italy. 
            
            HC, AFB and SK are supported by the Swiss National Science Foundation Grant 200021L\_189180 for STIX.
            L.A.H is supported by an ESA Research Fellowship.
            
\end{acknowledgements}

\bibliographystyle{aa} 
\bibliography{45293corr.bib} 

\end{document}